**Comment on "Proposed model for calculating the standard formation enthalpy of binary transition-metal systems"** [Appl. Phys. Lett. 81, 1219, (2002)].

O. Coreño Alonso[a*].

[a] Universidad de Guanajuato, Juárez # 77 Col. Centro, C.P. 36,000, Guanajuato, Gto. México.

*ocoreno@ugto.mx.

Miedema et al.[1] derived a method to predict thermodynamic properties of alloys, such as the heat of solution of metal B in metal A, and the standard formation enthalpy. In the case of the enthalpy of solution of *B* in *A*, at infinite dilution, it was expressed as:

$$H_{B\ in\ A}^{sol} = \frac{V_B^{2/3}}{\left(n_{ws}^{-1/3}\right)_{av}}\left[-P(\Delta\phi^*)^2 + Q\left(\Delta n_{ws}^{1/3}\right)^2\right] \quad (1)$$

Where *P* and *Q* are two empirical constants, $\phi^*$ is the chemical potential for electronic charge, and $n_{ws}$ is the electron density at the boundary of the Wigner-Seitz atomic cell.

Zhang and Liu[2] introduced a prefactor *S(C)* in equation (1) to consider the effect of the different atomic sizes of the atoms in the alloy. *S(C)* was defined as:

$$S(c) = 1 - S_v(c) \quad (2)$$

where $S_v(C)$, the influential factor, is the ratio between the difference of the surface area, and the average area of the dissimilar atoms[2]:

$$S_v(c) = \frac{C_B^S\left|V_A^{2/3} - V_B^{2/3}\right|}{C_A^S V_A^{2/3} + C_B^S V_B^{2/3}} \quad (3)$$

Then, *S(C)* can be expressed as

$$S(c) = 1 - \frac{C_B^S\left|V_A^{2/3} - V_B^{2/3}\right|}{C_A^S V_A^{2/3} + C_B^S V_B^{2/3}} \quad (4)$$

Where *c* is the alloy composition. $V_A^{2/3}$ and $V_B^{2/3}$ are the surface area of solute and solvent atoms, respectively, and $C_A^S$ and $C_B^S$ are the surface concentration of solute and solvent atoms, respectively. Surface concentrations are given by

$$C_A^S = C_A\, V_A^{2/3} / \left(C_A\, V_A^{2/3} + C_B\, V_B^{2/3}\right) \quad (5)$$

$$C_B^S = C_B\, V_B^{2/3} / \left(C_A\, V_A^{2/3} + C_B\, V_B^{2/3}\right) \quad (6)$$

After introducing the prefactor *S(C)*, Zhang and Liu[2] expressed equation (1) as:

$$\Delta H\ {}^{sol}_{B\ in\ A} = \frac{S(c)V_B^{2/3}}{\left(n_{ws}^{-1/3}\right)_{av}}\left[-P(\Delta\phi^*)^2 + Q\left(\Delta n_{ws}^{1/3}\right)^2\right] \tag{7}$$

And the standard formation enthalpy was given by

$$\Delta H_f^0 = f(c)\frac{S(c)V_B^{2/3}}{\left(n_{ws}^{-1/3}\right)_{av}}\left[-P(\Delta\phi^*)^2 + Q\left(\Delta n_{ws}^{1/3}\right)^2\right] \tag{8}$$

For intermetallic compounds,

$$f(c) = C_A^S[1 + 8(C_A^S C_B^S)^2] \tag{9}$$

$\Delta H_f^0$ values were calculated using equation (8) to compare them with the values reported by Zhang and Liu[2], Table I, column II. Only for $Ag_2Sc$, AgSc, and $Ag_{51}La_{14}$, the reported values, and those calculated in this work, Table I, column III, are equal. To check that correct values of $V_B^{2/3}, \phi^*,$ and $n_{ws}^{1/3}$ were used in this comment, $H_f^0$ values were calculated using the Miedema theory. The results are shown in Table I, column IV. Good agreement exists between these values, and those previously reported[2], Table I, column V. So, differences between the reported and calculated $\Delta H_f^0$ values must be caused by disagreement in *S(c)* values.

To find the cause of such differences, equation (10) is used instead of equation (6) to calculate $\Delta H_f^0$. Equation 10 follows the original the definition[2], expressed in words, of $S_v(c)$ .

$$S'(c) = 1 - S_v(c) = 1 - \frac{C_B\left|V_A^{2/3} - V_B^{2/3}\right|}{C_A\, V_A^{2/3} + C_B\, V_B^{2/3}} \tag{10}$$

Prefactor *S´(c)* values calculated for $Cu_4Sc$ and $CuSc_4$, using expression (10), are 0.8856 and 0.9148, respectively. They are in good agreement with *S(c)* reported[3] values, 0.8856 and 0.9149. In Table I, *S´(C)* values calculated using expression (10) are presented for all compounds reported by Zhang and Liu[2]. For these compounds, $\Delta H_f^0$ values were calculated by applying equation (8), using prefactor S´(c), Table I, column VI. $\Delta H_f^0$ values calculated using S´(c) are equal to, or close to those reported in the published letter[2]. Small differences could be due to different rounding procedures used therein, and in this work. On the other hand, $\Delta H_f^0$ values calculated using *S(c)* differ from the reported values[2]. Thus, according to the previous discussion, very possibly equation (6) in the reported letter[2] should be expressed as equation (10) above.

On the other hand, an analysis of equation (10), shows that *S´(C)* can be made independent of the difference in surface area, as shown below:

- If $V_A^{2/3} > V_B^{2/3}$, then $\left|V_A^{2/3} - V_B^{2/3}\right| = V_A^{2/3} - V_B^{2/3}$, and

$$S'(c) = 1 - \frac{C_B\left(V_A^{2/3} - V_B^{2/3}\right)}{C_A V_A^{2/3} + C_B V_B^{2/3}} = \frac{2\left(C_A V_A^{2/3} + C_B V_B^{2/3}\right) - V_A^{2/3}}{C_A V_A^{2/3} + C_B V_B^{2/3}}$$

$$S'(c) = 2 - \frac{V_A^{2/3}}{C_A V_A^{2/3} + C_B V_B^{2/3}} \qquad (11)$$

- If $V_A^{2/3} < V_B^{2/3}$, then $\left|V_A^{2/3} - V_B^{2/3}\right| = V_B^{2/3} - V_A^{2/3}$, and

$$S'(c) = 1 - \frac{C_B\left(V_B^{2/3} - V_A^{2/3}\right)}{C_A V_A^{2/3} + C_B V_B^{2/3}} = \frac{C_A V_A^{2/3} + C_B V_A^{2/3}}{C_A V_A^{2/3} + C_B V_B^{2/3}}$$

$$S'(c) = \frac{V_A^{2/3}}{C_A V_A^{2/3} + C_B V_B^{2/3}} \qquad (12)$$

for *S´(c)*, the physical meaning of constant 2 in (11) is unclear, as it is the ratio between solvent surface to average area of dissimilar atoms, in equations (11) and (12). Similar observations can be done if equation (4) for *S(c)* is analyzed, but in this case, area averaged using surface concentrations appears in the denominator of the corresponding equations. Then, the difference in size of the atoms, expressed as a difference in surface area, the justification stated by Zhang and Liu[2], for the introduction of S(c), is no longer present after the analysis.

TABLE I. $\Delta H_f^0$ (kJ/mol) reported in Ref. 2, recalculated according to original Eq. (8) of Rer. 2, recalculated according to Miedema theory, reported in Ref. 2, calculated using prefactor *S´(C),* prefactors *S(c),* and *S´(C).*

| Alloy system | Ref. 2 *II* | *Recalculated Equation (8) in Ref. 2. III* | Miedema theory, recalculated IV | Miedema[2] theory V | Miedema modified with *S´(c)* *VI* | *S (C)* | *S´ (c)* |
|---|---|---|---|---|---|---|---|
| $Cu_4Sc$ | -21 | -20 | -24 | -24 | -21 | 0.84153 | 0.88559 |
| $Cu_2Sc$ | -28 | -26 | -34 | -34 | -28 | 0.77423 | 0.82283 |
| CuSc | -26 | -24 | -34 | -34 | -26 | 0.71336 | 0.75587 |
| $Ag_2Sc$ | -34 | -34 | -38 | -38 | -34 | 0.89781 | 0.91178 |
| AgSc | -36 | -36 | -41 | -42 | -36 | 0.85946 | 0.87327 |
| AuSc | -95 | -94 | -109 | -109 | -95 | 0.85696 | 0.87118 |
| $Cu_4Y$ | -20 | -18 | -24 | -24 | -20 | 0.75406 | 0.83559 |
| $Cu_2Y$ | -24 | -21 | -32 | -32 | -24 | 0.67119 | 0.75305 |
| CuY | -20 | -18 | -30 | -30 | -20 | 0.60457 | 0.67029 |
| $Ag_{51}Y_{14}$ | -25 | -26 | -30 | -28 | -27 | 0.85758 | 0.89321 |
| AgY | -33 | -31 | -42 | -42 | -33 | 0.74747 | 0.78275 |
| AuY | -81 | -77 | -104 | -104 | -81 | 0.74492 | 0.78073 |
| $Cu_6La$ | -14 | -13 | -17 | -17 | -14 | 0.76578 | 0.85818 |
| $Cu_2La$ | -21 | -18 | -29 | -30 | -21 | 0.62493 | 0.72172 |
| $Ag_{51}La_{14}$ | -25 | -25 | -31 | -29 | -27 | 0.82039 | 0.87050 |
| AgLa | -30 | -28 | -40 | -41 | -30 | 0.69736 | 0.74331 |
| AuLa | -75 | -70 | -100 | -101 | -74 | 0.69483 | 0.74132 |

## AVAILABILITY OF DATA

Data sharing is not applicable to this article as no new data were created or analyzed in this study